\documentclass[12pt]{article}
\usepackage{graphicx}
\usepackage{color}
\usepackage{cite}
\def \beq{\begin{equation}}
\def \eeq{\end{equation}}
\def\bea{\begin{eqnarray}}
\def\eea{\end{eqnarray}}

\begin{document}
\setcounter{footnote}{1}
\rightline{EFI 15-12}
\rightline{arXiv:1502.01704}
\vskip1.5cm

\centerline{\large \bf LOW-ENERGY PHOTON PRODUCTION}
\centerline{\large \bf IN NEUTRINO NEUTRAL-CURRENT INTERACTIONS}
\bigskip

\centerline{Jonathan L. Rosner\footnote{{\tt rosner@hep.uchicago.edu}}}
\centerline{\it Enrico Fermi Institute and Department of Physics}
\centerline{\it University of Chicago, 5620 S. Ellis Avenue, Chicago, IL
60637, USA}
\bigskip

\begin{center}
ABSTRACT
\end{center}
\begin{quote}
The search for $\nu_\mu \to \nu_e$ oscillations by the MiniBooNE Collaboration
at Fermilab has revealed a low-energy signal which could be due either to
electrons produced by $\nu_e$ or photons produced by the interaction
of the weak neutral current on the target nucleus.  One contribution to the
latter is a Wess-Zumino-Witten anomaly leading to a term in the Lagrangian
proportional to $\epsilon^{\mu \nu \kappa \lambda} Z_\mu \omega_\nu
F_{\kappa \lambda}$.  This term is normalized with the help of the known rates
for the processes $f_1 \to \rho \gamma$ and $\tau \to \nu_\tau a_1$.  A rate
of about 1/4 of that employed in several previous estimates is obtained.  As
the anomaly term had already been found to play a subdominant role in
photon production (e.g., in comparison with $\Delta$ excitation and decay), the
present estimate reduces its strength even further.
\end{quote}
\smallskip

\leftline{PACS codes: 12.15.Mm, 13.15.+g, 13.35.Dx, 14.60.Lm}
\bigskip

\section{INTRODUCTION \label{sec:int}}
The search for $\nu_\mu \to \nu_e$ oscillations by the MiniBooNE Collaboration
at Fermilab \cite{MB} has revealed a low-energy signal which could be due
either to electrons produced by $\nu_e$ or photons produced by the coherent
interaction of the weak neutral current on the target nucleus \cite{Gershtein:%
1980wu,Rein:1981ys,HHH}.  One proposed source of the latter is an interaction
between the $Z$, $\omega$ meson, and photon \cite{DH} due to a Wess-Zumino-%
Witten anomaly \cite{WZ,W,Harada:2011xx}, whose strength has been calculated
in Refs.\ \cite{HS,SS,DGH}.  It was concluded in Ref.\ \cite{Hill:2009ek,%
Hill:2010zy} that the anomaly contribution was not enough to account for a
photon signal.  Although neutral-current nucleon excitation followed by photon
emission was originally suggested as a source of the signal, it was found
insufficient as well in Refs.\ \cite{Serot:2012rd,Zhang:2012aka,Zhang:2012xi,%
Zhang:2012xn,Wang:2013wva,Wang:2014nat}.  A comprehensive review of neutrino-%
induced quasi-elastic scattering and single photon production is given in a
workshop summary \cite{Garvey:2014exa}.

The imminent operation of the MicroBooNE Experiment at Fermilab \cite{MuB}
will be able to distinguish final-state photons from electrons.  Hence it
is timely to present an independent estimate of the strength of the
anomaly-mediated interaction.  In this paper we perform such an estimate based
on dominance of the $Z$--$\omega$--$\gamma$ interaction by the $a_1$ pole
in the neutral current.  The $a_1$ decay constant is obtained from the
observed rate for $\tau \to a_1 \nu_\tau$, while the $a_1$--$\omega$--$\gamma$
coupling is obtained from the observed decay rate for $f_1 \to \gamma \rho$,
which involves a coupling constant identical to $g_{a_1\omega\gamma}$ if $f_1$
contains only nonstrange quarks.  This interaction was overlooked in an
otherwise successful description of light meson radiative decays based on the
quark model \cite{BR,JR81}.

In Sec.\ \ref{sec:anom} we review the consequence for the
MiniBooNE experiment of the assumed $Z$--$\omega$--$\gamma$ interaction. We
then (Sec.\ \ref{sec:adom}) derive the consequence of assuming $a_1$ dominance
of the weak neutral current.  The $a_1$ decay constant which arises in this
derivation is evaluated with the help of the rate for $\tau \to a_1 \nu_\tau$
in Sec.\ \ref{sec:adec}, while the decay rate for $f_1 \to \gamma \rho$ is
employed to evaluate the $a_1$--$\omega$--$\gamma$ coupling in Sec.\
\ref{sec:fdec}.  Section \ref{sec:apred} contains predictions for the rates
for $a_1^0 \to \omega \gamma$ and $a_1 \to \rho \gamma$ (all charge states).
We sum up in Sec.\ \ref{sec:summ}.

\section{NEW INTERACTION AND MINIBOONE \label{sec:anom}}

The MiniBooNE experiment \cite{MB} at Fermilab was conceived to check
a signal for $\bar \nu_\mu \to \bar \nu_e$ oscillation observed at the LSND
detector \cite{LSND} operating at Los Alamos National Laboratory.  The
oscillation signature would be the appearance of electrons.  Initially 
signals were restricted to an electromagnetic energy deposit greater than
475 MeV.  An excess of events below this cutoff was observed, attributable
either to electrons or to photons.

A possible source of photons in this experiment was identified by J. A. Harvey,
C. T. Hill, and R. J. Hill \cite{HHH}.  A Wess-Zumino-Witten anomaly
\cite{WZ,W} gives rise to a term
\beq \label{eqn:dl}
\delta{\cal L} = \frac{N_c}{48 \pi^2} \frac{e g_\omega g_2}{\cos \theta_W}
\epsilon_{\mu\nu\rho\sigma}\omega^\mu Z^\nu F^{\rho\sigma}~.
\eeq
Here $N_c = 3$ is the number of quark colors, $e = \sqrt{4 \pi \alpha} =
0.3028$ is the proton charge, $g_\omega$ is a coupling constant of the $\omega$
meson to baryon number whose value needs to be specified, $g_2$ is the
electroweak SU(2) coupling constant, $\theta_W$ is the electroweak mixing
angle, and $F^{\rho\sigma}$ is the photon field-strength tensor.  The
contribution of this term to the coherent process $\nu A \to \nu \gamma A$,
where $A$ denotes a nucleus of atomic number $A$, is illustrated in Fig.\ 1(a).
The induced decay $f_1 \to \gamma \rho$ is shown in Fig.\ 1(b), while a
related WZW anomaly \cite{WZ,W} leads to a $K \bar K \omega$ coupling
responsible for $K_L \to K_S$ coherent regeneration [Fig.\ 1(c)]
\cite{Cabibbo:1966zza,Roehrig:1977db}.
\begin{figure}
\begin{center}
\includegraphics[width=0.96\textwidth]{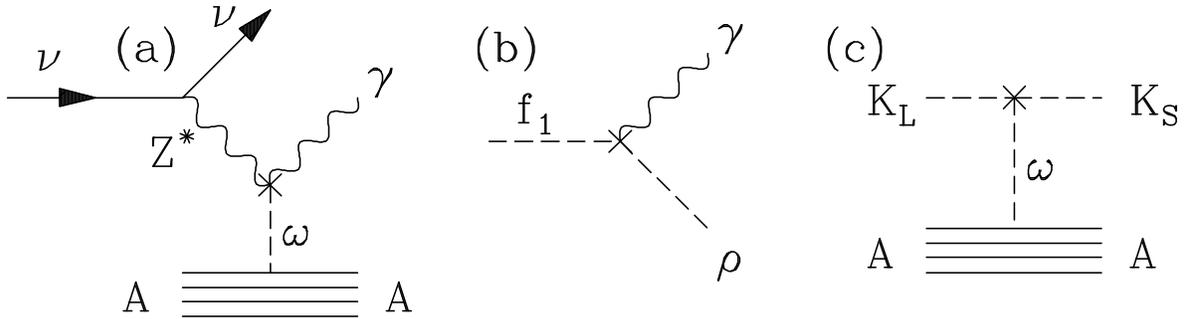}
\end{center}
\caption{Processes sensitive to WZW anomalies:  (a) Coherent reaction $\nu A
\to \nu \gamma A$ on a nucleus of atomic number $A$ ; (b) Induced $f_1 \to
\gamma \rho$ decay; (3) Coherent $K_L \to K_S$ regeneration on a nucleus of
atomic number $A$.
\label{fig:anom}}
\end{figure}

The term (\ref{eqn:dl}) leads to a cross section per nucleon $N$ in the
zero-recoil limit \cite{Gershtein:1980wu,HHH}:
\beq \label{eqn:sig}
\sigma(\nu N \to \nu \gamma N) = \frac{\alpha g_\omega^4 G_F^2}{480 \pi^6
m_\omega^4}E_\nu^6
 = 2.6 \times 10^{-41}~E^6_\nu({\rm GeV})(g_\omega/10.0)^4 {\rm cm}^2~.
\eeq
in the limit of $E_\nu$ below a few hundred MeV.  The $E_\nu^6$ behavior can
only be valid below threshold for nucleon excitation; contributions of
Compton-like scattering and $\Delta$ excitation are more important in this
regime \cite{Hill:2009ek,Hill:2010zy}.  (For $E_\nu \simeq 20$ GeV, upper
limits on neutral-current photon production contradict (\ref{eqn:sig})
\cite{Kullenberg:2011rd}.) For production on a nucleus of atomic number $A$,
the right-hand side of Eq.\ (\ref{eqn:sig}) is to be multiplied by a
factor less than $A$ \cite{Hill:2009ek}.

Ref.\ \cite{HHH} quoted considerable uncertainty in the value of $g_\omega$ but
noted that the nominal value $g_\omega = 10$ agreed roughly with the MiniBooNE
signal interpreted as photons.  This was found \cite{Hill:2009ek,%
Hill:2010zy} to be an overestimate, as a result of form factor and recoil
effects, with $\Delta$ excitation and decay providing a dominant photon source,
but the nominal value $g_\omega = 10$ was retained.  Corrections for efficiency
\cite{Zhang:2012xn,Wang:2014nat} further reduced the expected signal.
We shall use $a_1$
dominance of the weak neutral current to find an independent extimate of
$g_\omega$, finding a value close to 7.  This strengthens the conclusions of
Refs.\ \cite{HS,SS,DGH,Hill:2009ek,Hill:2010zy,Serot:2012rd,Zhang:2012aka,%
Zhang:2012xn,Wang:2013wva,Wang:2014nat} that if the MiniBooNE signal is indeed
found to be photons, the WZW anomaly is unlikely to be their dominant source.

\section{$a_1$ DOMINANCE OF NEUTRAL CURRENT \label{sec:adom}}

In the contribution (\ref{eqn:dl}) to the Lagrangian, the weak neutral current
carried by the $Z$ may be viewed as dominated by the $a_1(1260)$ meson.  The
matrix element between $Z$ and $a_1$ may be written as $m_a f_a$, where
\cite{PDG2014} $m_a = 1230 \pm 40$ MeV and the neutral $a_1$ decay constant is
to be determined.  The interaction between a $Z$ and a fermion-antifermion pair
$f \bar f$ is
\beq
{\cal L}{Z f \bar f} = \frac{g_2}{\cos \theta_W} \bar f \gamma_\mu Z^\mu
[(1-\gamma_5) a_L + (1+\gamma_5) a_R ]f~,
\eeq
where the coupling constants $a_L$ and $a_R$ are listed in Table \ref{tab:cc}.

\begin{table}
\caption{Coupling constants of $f \bar f$ to the $Z$.  Here $x_W \equiv \sin^2
\theta_W$.
\label{tab:cc}}
\begin{center}
\begin{tabular}{c c c} \hline \hline
 $f\bar f$ pair  &     $a_L$     &   $a_R$    \\ \hline
 $\nu \bar \nu$  & $\frac{1}{4}$ &     0      \\
$\ell \bar \ell$ & $\frac{1}{2} \left( -\frac{1}{2} + x_W \right)$ &
 $\frac{1}{2} x_W$ \\
 $u \bar u$ & $\frac{1}{2} \left( \frac{1}{2} - \frac{2}{3} x_W \right)$ &
 $\frac{1}{2} \left( -\frac{2}{3} x_W \right)$\\
 $d \bar d$ & $\frac{1}{2} \left( -\frac{1}{2} + \frac{1}{3} x_W \right)$ &
 $\frac{1}{2} \left( \frac{1}{3} x_W \right)$ \\
\hline \hline
\end{tabular}
\end{center}
\end{table}

The axial vector coupling is then
\beq
{\cal L}^{\rm axial}_{Z f \bar f} = \frac{g_2}{\cos \theta_W} \bar f \gamma_mu 
\gamma_5 Z^\mu (a_R - a_L) f~,
\eeq
where
\beq
(a_R - a_L)(u \bar u) = -\frac{1}{4}~,~~(a_R - a_L)(d \bar d) = \frac{1}{4}~.
\eeq
The quark content of $a_1$ is $(d \bar d - u \bar u)/\sqrt{2}$, so the
$Z$--$a_1^0$ coupling may be written as
\beq
g_{Z {a_1}^0} = \frac{g_2 f_{a} m_a}{2 \sqrt{2} \cos \theta_W}~.
\eeq

Assuming $a_1$ dominance of the weak neutral current, Eq.\ (\ref{eqn:dl})
then may be written as
\beq \label{eqn:dla}
\delta{\cal L} = \frac{f_a}{2 \sqrt{2} m_a}\epsilon_{\mu\nu\rho\sigma}
\omega^\mu Z^\nu F^{\rho\sigma}~g_{a \omega \gamma}~.
\eeq
Equating coefficients of equal terms in Eqs.\ (\ref{eqn:dl}) and 
(\ref{eqn:dla}) and taking $N_c = 3$, we find
\beq \label{eqn:aomg}
g_{a \omega \gamma} = \frac{e g_\omega}{4 \sqrt{2}\pi^2}~\frac{m_a}{f_a}~.
\eeq
We shall evaluate $f_a$ in the next Section.

\section{EVALUATION OF $f_{a_1}$ \label{sec:adec}}

The decays $\tau^- \to \pi^- \nu_\tau$ and $\tau^- \to \rho^- \nu_\tau$
are described by simple expressions involving the pion and rho decay
constants, respectively (see, e.g., Ref.\ \cite{JR90}.)  The corresponding
expression for the decay $\tau^- \to a_1^- \nu_\tau$, in terms of the decay
constant $f_a$ linking the $Z$ and the neutral $a_1$, is
\beq
\Gamma(\tau^- \to a_1^- \nu_\tau) = \frac{G_F^2 m_\tau^3 f_a^2}{8 \pi}
|V_{ud}|^2\left( 1 + \frac{2m_a^2}{m_\tau^2} \right) \left(1 - \frac{m_a^2}
{m_\tau^2} \right)^2~.
\eeq
The Particle data Group does not give a branching fraction for this decay.
However, assuming that the quoted branching fractions \cite{PDG2014}
\beq
{\cal B}(\tau^- \to \pi^- \pi^0 \pi^0 \nu_\tau) = (9.30 \pm 0.11)\%~,~~
{\cal B}(\tau^- \to \pi^- \pi^+ \pi^- \nu_\tau) = (8.99 \pm 0.06)\%
\eeq
are dominated by the $a_1$, one obtains ${\cal B}(\tau^- \to a_1^- \nu_\tau) =
(18.29 \pm 0.13)\%$.  Using the values $G_F=1.16638 \times 10^{-5}$ GeV$^{-2}$,
$m_\tau = 1776.82 \pm 0.16$ MeV, $|V_{ud}| = 0.97425 \pm 0.00022$, $m_a =
1230 \pm 40$ MeV, and $\tau_\tau = 290.3 \pm 0.5$ fs quoted in \cite{PDG2014},
we find a decay constant $f_a = (164.6 \pm 7.3)$ MeV, where the error is
dominated by uncertainty in $m_a$.  In our convention $f_a$ refers to the {\it
neutral} current, whereas most authors quote a value which in our notation
would be $\sqrt{2} f_a$.  Our value is consistent with several others obtained
theoretically or extracted from data \cite{Wingate:1995hy,Dhir:2013zia}.  The
resulting decay constant may now be used in Eq.\ (\ref{eqn:aomg}) to obtain
the result $g_{a\omega\gamma} = (0.0405 \pm 0.0005) g_\omega$.

\section{EVALUATION OF $g_{a\omega\gamma}$ USING $f_1 \to \gamma \rho$ RATE
\label{sec:fdec}}

The coupling constants $g_{a\omega\gamma}$ and $g_{f \rho \gamma}$ both
involve the isovector photon coupling to an isovector and isosinglet, and
are equal by U(3) symmetry as long as $f_1$ contains no strange quarks:
$g_{f \rho \gamma} = g_{a\omega\gamma}$.  The rate for $f_1 \to \gamma \rho$
is given (see also Appendix C of Ref.\ \cite{DGH}) by 
\beq \label{eqn:frat}
\Gamma(f_1 \to \gamma \rho) = \frac{g^2_{f\rho\gamma}}{3\pi} \frac{E^3_\gamma}
{m_\rho^2} \left( 1 + \frac{m_\rho^2}{m_f^2} \right) = (26.6 \pm 0.7)
 g_\omega^2~{\rm keV}~,
\eeq
where we have used $m_\rho = (775.25 \pm 0.25)$ MeV, $m_f = (1281.9 \pm 0.5)$
MeV, and $E_\gamma = (m_f^2 - m_\rho^2)/(2 m_f) = (406.5 \pm 0.4)$ MeV.
The first and second terms in large parentheses correspond to longitudinal and
transverse $\rho$ polarizations, respectively, so longitudinal $\rho$
polarization is dominant \cite{BR,Amelin:1994ii}, in contradiction to the
result found in Ref.\ \cite{Coffman:1989nk}.

The experimental partial width for $f_1 \to \gamma \rho$ is the product of
the total $f_1$ width and the corresponding branching fraction
\cite{PDG2014}:
\beq \label{eqn:fw}
\Gamma(f_1 \to \gamma \rho) = \Gamma(f_1) {\cal B}(f_1 \to \rho \gamma) =
(24.2 \pm 1.1)~{\rm MeV})(0.055 \pm 0.013) = (1.33 \pm 0.32)~{\rm MeV}~.
\eeq
When combined with Eq.\ (\ref{eqn:frat}) this yields $g_\omega = 7.07 \pm 0.86$,
where the error is dominated by the experimental error in Eq.\ (\ref{eqn:fw}).
This value is lower than the nominal one of 10 taken in \cite{HHH,Hill:2009ek},
and leads to an anomaly contribution only about 1/4 of that previously
estimated, thanks to the quartic power of $g_\omega$ in Eq.\ (\ref{eqn:sig}).

The systematic errors that we are able to identify tend to
decrease $g_\omega$ by a modest amount.  We have taken ${\cal B}(\tau \to
a_1 \nu_\tau)$ to be as large as possible when ascribing all the $3\pi
\nu_\tau$ decays to $a_1$.  If ${\cal B}(\tau^- \to a^-_1 \nu_\tau)$ is
smaller, $f_a$ is smaller, the coefficient of $g_\omega$ is larger in Eq.\
(\ref{eqn:aomg}), so $g_\omega$ is smaller.  We have also assumed the anomaly
to fully account for the $f_1 \to \rho \gamma$ decay rate, whereas a small
quark model contribution of 150 keV was predicted in Ref.\ \cite{BR}.  It is
not clear whether the decay amplitude for longitudinal $\rho$ production
predicted in Ref.\ \cite{BR} should be added coherently to that predicted here.

\section{RATES FOR $a^0_1 \to \gamma \omega$ AND $a_1 \to \gamma
\rho$ \label{sec:apred}}

The decay $f_1 \to \gamma \rho$ is related by U(3) symmetry to the
decays $a_1 \to \gamma \omega$ and $a_1 \to \gamma \rho$:
\beq
g^2_{a \omega \gamma} = 9 g^2_{a \rho \gamma} = g^2_{f \rho \gamma} =
0.082 \pm 0.020~,
\eeq
where we have used the experimental value (\ref{eqn:fw}) in the expression
(\ref{eqn:frat}) for the $f_1 \to \gamma \rho$ rate.  The corresponding
formulae for the $a_1$ radiative decay widths are
\bea \label{eqn:aomrat}
\Gamma(a^0_1 \to \gamma \omega) & = & \frac{g^2_{a \omega \gamma}}{3\pi}
 \frac{E^3_{\gamma}}
 {m_\omega^2} \left( 1 + \frac{m_\omega^2}{m_a^2} \right)~, \\
\Gamma(a_1 \to \gamma \rho) & = & \frac{g^2_{a \rho \gamma}}{3\pi}
 \frac{E^3_{\gamma}}{m_\rho^2} \left( 1 + \frac{m_\rho^2}{m_a^2} \right)~,
\eea
where the photon energies, predicted rates, and range of predicted branching
fractions (using $\Gamma_{\rm tot}(a_1) = 250$ to 600 MeV \cite{PDG2014})
are shown in Table \ref{tab:apreds}.  We have used $m_\omega=(782.65\pm0.12)$
MeV.  The expression (15) holds for all $a_1$ charge states.
\begin{table}
\caption{Photon energies, predicted decay rates, and range of predicted
branching fractions for the decays $a_1 \to \gamma \omega$ and $a_1 \to \gamma
\rho$.
\label{tab:apreds}}
\begin{center}
\begin{tabular}{c c c c} \hline \hline
Final state $f$ & $E_\gamma$ (MeV) & $\Gamma_f$ & ${\cal B}_f$ \\ \hline
$\gamma \omega$ & 366 & $(0.98 \pm 0.24)$ MeV & (1.2--4.9)$\times 10^{-3}$ \\
$\gamma \rho$   & 371 & $(115 \pm 28)$ keV & (1.4--5.7)$\times 10^{-4}$ \\
\hline \hline
\end{tabular}
\end{center}
\end{table}

The branching fractions in Table \ref{tab:apreds} are quite small because of
the large $a_1$ total width.  Nevertheless, it may be possible to see the
decay $a_1^- \to \gamma\rho^-$ in the final state of $\tau^- \to a^-_1
\nu_\tau$.  The subprocess $a^0_1 \to \gamma \omega$ may be observable through
coherent photoproduction of $a^0_1$ on a heavy nucleus $a=A$:  $\gamma A \to
a_1^0 A \to \pi^+ \pi^- \pi^0 A$, proceeding via $\omega$ exchange.

\section{SUMMARY \label{sec:summ}}

A neutral-current interaction based on the $Z$--$\omega$--$\gamma$ interaction
depicted in Fig.\ 1(a) is predicted \cite{HHH} on the basis of a
Wess-Zumino-Witten anomaly \cite{WZ,W}.  This interaction is calibrated with
the help of the decay $\tau \to a_1 \nu$.  It leads to a prediction for the
low-energy signal in the MiniBooNE experiment \cite{MB}, if interpreted in
terms of photons rather than electrons or positrons, which is about a fourth of
that previously estimated, which already was below the needed magnitude.
(One proposed source of photons is the decay of a quasi-sterile neutrino with
mass between 40 and 80 MeV \cite{Gninenko:2009ks,Gninenko:2010pr}.)

For the future one looks forward to tests of the predicted rates for $a_1 \to
\gamma \omega$ and $a_1 \to \gamma \rho$, to a more precise estimate of $f_a$,
and to an experimental distinction between electrons or positrons
and photons in the final state studied by MiniBooNE.  The MicroBooNE experiment
\cite{MuB}, soon to begin operation at Fermilab, should resolve the question.

\section*{ACKNOWLEDGMENTS}
I thank S. Domokos, J. Harvey, R. Hill, S. Gninenko, M. Khlopov, P. Langacker,
W. Louis, W. A. Mann, L. A. Ruso, and A. Weinstein for helpful discussions.
This work was supported in part by the U. S. Department of Energy under Grant
No.\ DE-FG02-13ER41598 and in part by funds from the Physics Department of the
University of Chicago.

\end{document}